\documentclass[12pt,preprint]{aastex}

\slugcomment{To Appear in The Astrophysical Journal}

\shorttitle{The Debris System in HD 12039}
\shortauthors{Hines et al.}

\begin{document}


\title{The Formation and Evolution of Planetary Systems (FEPS):\\
Discovery of an Unusual Debris System Associated with HD~12039}

\author{Dean C. Hines\altaffilmark{1}, 
Dana E. Backman\altaffilmark{2}, 
Jeroen Bouwman\altaffilmark{3}, 
Lynne A. Hillenbrand\altaffilmark{4},
John M. Carpenter\altaffilmark{4},
Michael R. Meyer\altaffilmark{5},
Jinyoung Serena Kim\altaffilmark{5},
Murray D. Silverstone\altaffilmark{5},
Jens Rodmann\altaffilmark{3},
Sebastian Wolf\altaffilmark{3}, 
Eric E. Mamajek\altaffilmark{6},
Timothy Y. Brooke\altaffilmark{4}, 
Deborah L. Padgett\altaffilmark{7},
Thomas Henning\altaffilmark{3},
Amaya Moro-Mart{\'{\i}}n\altaffilmark{8},
E. Stobie\altaffilmark{5},
Karl D. Gordon\altaffilmark{5},
J.E. Morrison,\altaffilmark{5},
J. Muzerolle\altaffilmark{5},
and
K.Y.L. Su\altaffilmark{5}
}
\altaffiltext{1}{Space Science Institute, 4750 Walnut Street,
Suite 205 Boulder, CO 80301}

\altaffiltext{2}{SOFIA, MS 211-3, NASA-Ames, Moffet Field, CA 94035-1000}

\altaffiltext{3}{Max-Planck-Institut fu$\:$r Astronomie, D-69117
Heidelberg, Germany.}

\altaffiltext{4}{Astronomy, California Institute of Technology,
Pasadena, CA 91125.} 

\altaffiltext{5}{Steward Observatory, The
University of Arizona, 933 N. Cherry Ave., Tucson, AZ 85721}

\altaffiltext{6}{Harvard-Smithsonian Center for Astrophysics, Cambridge, MA}

\altaffiltext{7}{{\it Spitzer} Science Center (SSC), California
Institute of Technology, Pasadena, CA 91125.}

\altaffiltext{8}{Princeton University, Princeton, NJ 08540}


\begin{abstract}
We report the discovery of a debris system associated with the $\sim
30$ Myr old G3/5V star HD~12039 using {\it Spitzer Space Telescope}
observations from 3.6 -- 160$\mu$m.  An observed infrared excess
(L$_{\rm IR}$/L$_{\ast} = 1\times10^{-4}$) above the expected
photosphere for $\lambda \gtrsim 14\mu$m is fit by thermally emitting
material with a color temperature of T$\sim 110$ K, warmer than the
majority of debris disks identified to date around Sun-like stars.
The object is not detected at 70$\mu$m with a 3$\sigma$ upper limit 6
times the expected photospheric flux.
The spectrum of the infrared excess can be explained by warm,
optically thin material comprised of blackbody-like grains of size
$\gtrsim 7 \mu$m that reside in a belt orbiting the star at 4--6 AU.
An alternate model dominated by smaller grains, near the blow-out size
$a\sim 0.5\mu$m, located at 30-40AU is also possible, but requires the
dust to have been produced recently since such small grains will be
expelled from the system by radiation pressure in $\sim$ few $\times
10^{2}$yrs.

\end{abstract}


\section{Introduction}

Dust in the inner Solar System can be seen via scattered light as the
zodiacal glow visible to the naked eye near sunrise and sunset in dark
clear skies.  This dust is located within the terrestrial planet zone,
arising primarily from collisions among rocky debris (asteroids)
between Mars and Jupiter and secondarily from comet ejecta.  Dust
produced by collisions between remnant planetesimals is also expected
to exist within the Kuiper Belt beyond Neptune.  The region between 5
and 30 AU in our solar system is kept relatively free of dust due to
the dynamical action of the giant planets (Moro-Mart{\'{\i}}n \&
Malhotra 2002, 2003; Moro-Mart{\'{\i}}n, Wolf \& Malhotra 2005).
Comparable planetary systems around other stars might be expected to
produce similar dusty debris belts and gaps.

Examples of thermal emission from debris disks around fairly luminous
main-sequence stars, including the proto-type Vega (Aumann et al.
1984), were identified with {\it IRAS} (e.g., Walker \& Wolstencroft
1988; Backman \& Paresce 1993; Mannings \& Barlow 1998) and {\it ISO}
(e.g., Habing et al.  2001; Spangler et al.  2001; Laureijs et al.
2002).  Most of these systems are associated with A-type stars since
neither observatory had the sensitivity to detect the amount of
material associated with Vega around low-luminosity Sun-like stars at
distances beyond about 15-20 parsecs.  Most debris systems identified
to date are characterized by cool (T $\sim 50-90$K) disks, suggestive
of material at gas giant or Kuiper Belt distances from their parent
stars, r$_{\rm dust} \sim 10-100$ AU (see, e.g., Backman \& Paresce
1993; Decin et al.  2000, 2003; Lagrange, Backman \& Artymowicz 2000;
Zuckerman 2001 and references therein).

Since much of the zodiacal dust in our solar system arises from
collisions in the asteroid belt (e.g., Grogan et al.  2001; Nesvorn{\'
y} et al.  2002, 2003), we might expect to observe excess emission
from debris in a similar zone around other stars (Stern 1994; Zhang \&
Sigurdsson 2003).  However, only a handful of debris disks associated
with less luminous, Sun-like stars have been found to exhibit peak
infrared excess emission from warm dust (T $\sim 100-300$K) suggestive
of terrestrial zone material (r$_{\rm dust}\sim 2-5$ AU).
Herein, we report the discovery of another example that is associated
with HD~12039 (DK Cet, HIP 9141), a young ($\approx$30 Myr) G3/5V star
(Houk et al.  1988) at 42\,$\pm$\,2\,pc (Perryman 1997).  The infrared
excess was identified in
the initial enhanced data product release from our {\it Spitzer Space
Telescope} ({\it Spitzer}: Werner et al.  2004) legacy science program
titled FEPS (Formation and Evolution of Planetary Systems: Meyer et
al.  2004, 2005; Hines et al.  2004a).

Sections 2 and 3 briefly describe the FEPS program and the {\it
Spitzer} observations of HD~12039.  Section 4 describes our discovery
of the debris system associated with HD~12039, and presents some of
the properties of the star and its infrared excess.  We further
characterize and model the debris system in Section 5.  Section 6
explores the origin of the emitting material and how HD~12039 fits
into models of disk evolution and planet formation.  We also compare
HD~12039 to other warm debris systems associated with Sun-like stars.

\section{The FEPS Program and V1.1 Sample}

The FEPS {\it Spitzer} Legacy program encompasses observations of 328
Sun-like ($\sim$ 0.8--1.5 M$_{\odot}$) stars with ages ranging from
$\sim$ 3 Myr to 3 Gyr.  The survey enables us to examine the frequency
of stars that have circumstellar gas and dust, as well as to infer
some of the properties of detected dust (Meyer et al.  2004; Meyer et
al.  2005).
Our ultimate goal is to assess whether debris systems such as that
found in our solar system are common or rare around Sun--like stars in
the disk of the Milky Way.  FEPS uses all three {\it Spitzer} science
instruments to provide spectral coverage from 3.6$\mu$m to 70$\mu$m,
and includes $\lambda \sim 7 - 35 \mu$m low resolution IRS spectra.
We also obtain 160 $\mu$m observations for a subset of the FEPS stars
located in areas of low infrared cirrus background.  In addition, FEPS
provides models of the photospheric emission that have been fit to
available optical and near-infrared photometry for each star.

The initial FEPS data release contains $\approx 10\%$ (33 stars) of
our total sample.  A detailed description of the observing strategy,
data analysis and best-fit, Kurucz photospheric emission models for
these stars are provided in the FEPS Explanatory Supplement (Version
1.1: Hines et al.  2004a)\footnote{Subsequent to the release of V1.1,
the IRS spectra were reprocessed by an updated SSC pipeline that
significantly improved the calibration.  The MIPS Instrument Team also
released an improved data analysis pipeline that reduced the noise at
70$\mu$m.  These updated IRS and MIPS data are used herein and have
been released publicly in FEPS V2.0.}.  The {\it Spitzer} data and
documentation for the FEPS program are available directly from the
{\it Spitzer} Legacy Science
Archive\footnote{http://data.spitzer.caltech.edu/popular/feps/}.

The 33 stars were randomly chosen by the SSC for scheduling between UT
May 2004 and UT September 2004.  They cover a wide range of properties
within the selection criteria used for the entire FEPS program, and
therefore provide a good sub-sample from which to begin looking for
interesting objects.
In particular, these stars cover a broad range of effective
temperatures $5121K \le T_{\rm 33} \le 6227K$ (compared with $4299K \le T_{\rm
FEPS} \le 6769K$ for the entire sample),
have an age range $7.2 \lesssim$ log(age) $\lesssim 9.2$ [yrs], and
lie at distances $20 \le d \le 127$ pc.  FEPS stars with younger ages
6.2 $\le$ log(age) $\lesssim 7.2$ and greater distances $d \gtrsim
130$ pc are primarily cluster members and are less well represented by
these 33 stars.

FEPS results to date include: 1) the frequency of young, optically
thick disks detected with the IRAC instrument (Silverstone et al.
2005); 2) cool debris systems having excesses detected only at MIPS
70$\mu$m and therefore comparable to our own Kuiper Belt (Meyer et al.
2004; Kim et al.  2005): and 3) constraints on the amount of remnant
gas in the disk surrounding HD~105 (Hollenbach et al.  2005).  We have
also investigated the occurrence of debris disks in the 100 Myr old
Pleiades star cluster (Stauffer et al.  2005).  We focus the current
study on the Sun-like star HD~12039, which exhibits dust emission with
temperatures that are warmer than found in typical disks around other
stars.

\section{{\it Spitzer} Observations of HD~12039}

In this section we briefly review the observing strategy and data
reduction methods specifically for HD~12039.  Observations of the
other 32 stars were executed and processed in a similar manner.
Additional details are presented in the Explanatory Supplements for
the FEPS V1.1 and V2.0 data releases (Hines et al.  2004a, 2005), and
in Kim et al.  (2005), Bouwman et al.  (2005), and Silverstone et al
(2005) for MIPS, IRS and IRAC respectively.

Multi-band Imaging Photometer for {\it Spitzer} (MIPS: Rieke et al.
2004) observations of HD~12039 were obtained using the small field
photometry mode on UT 2004 July 11 at 24, 70 and 160$\mu$m.  Two
cycles of 3s Data Collection Events (DCEs) at 24$\mu$m, two cycles of
10s DCEs at 70$\mu$m, and four cycles of 10s DCEs at 160$\mu$m were
obtained.
After initial processing by the SSC S10.5.0 pipeline to provide
reconstructed pointing information, the MIPS data were further
processed using the MIPS Data Analysis Tool (DAT, ver.  2.9; Gordon et
al.  2004, 2005).  This includes the ``enhancer'' routines that
correct for distortion in individual images and combine the images
onto a sub-sampled mosaic.  Aperture photometry using IDP3 (version
2.9: Schneider \& Stobie 2002) was performed on the individual
24$\mu$m DCE images, and on the mosaic images at 70$\mu$m and
160$\mu$m.  We adopt 14\farcs7, 29\farcs7 and 48\farcs0 target
aperture radii at 24, 70, and 160$\mu$m, respectively.  Background
annuli from 29\farcs4 -- 41\farcs7, 39\farcs6 -- 79\farcs2 and
48\farcs0 -- 80\farcs0 were used for the three bands respectively.
The median background per pixel was scaled to the appropriate target
aperture size and subtracted from the summed target aperture flux for
each star.  The final flux in instrumental units was then corrected to
an infinite aperture (MIPS Data Handbook, V3.1, hereafter MDH3.1).

Random uncertainties in the background--subtracted estimates of
on--source flux were estimated from
the standard deviation of the mean of the multiple measurements for
the 24$\mu$m observations.  These ``internal'' uncertainties represent
the precision of our measurements independent of the absolute flux
calibration of {\it Spitzer}.
The ``internal'' uncertainties for the 70$\mu$m and 160$\mu$m
observations of HD~12039 were estimated by the {\it rms}
pixel-to-pixel dispersion inside the background annulus measured on
the mosaic image and scaled to the area of the target aperture.

Conversion to physical flux density units was performed with a simple
multiplication of the instrumental total flux by the MIPS calibration
factor as published currently in the MDH3.1.  The MIPS absolute
calibration uncertainties for the 24, 70, and 160 $\mu$m wavelength
bands are currently listed as $10\%$, 20\%, and 20\%, respectively in
the {\it Spitzer} Observer's Manual Version 4.6 (hereafter SOM V4.6),
and in the MDH3.1\footnote{The MDH3.1 states that the absolute
calibration at 24$\mu$m is ``better than $10\%$.''}.  However,
the median of the ratio of the measured 24$\mu$m flux densities to
Kurucz-model predicted 24$\mu$m flux densities for the ensemble of 33
FEPS stars is 0.98 with a standard deviation 0.05 (\S 4.2).  Assuming
that the stars do not have infrared excesses, this small dispersion
indicates that the MIPS 24$\mu$m absolute calibration is actually
accurate to $\sim 5\%$.  This uncertainty estimate is consistent with
24$\mu$m measurements of 69 F5-K5 (IV \& V) stars observed by the MIPS
GTO science team (Beichman et al.  2005a; Bryden et al.  2005).
Excluding the sole warm debris system star HD~69830, Bryden et al.
find that the mean MIPS 24$\mu$m to predicted 24$\mu$m flux density
ratio for the remaining 68 stars is 0.99 with a dispersion of 0.06, in
excellent agreement with our results.  We conclude that the
photometric calibration of our MIPS 24$\mu$m photometry relative to
the best fit Kurucz models is 5-6\%.  Since we do not detect the
photospheres at 70 and 160$\mu$m, we cannot independently derive
uncertainties relative to Kurucz models.  Therefore we adopt the
SSC-published 20\% absolute uncertainty for both the MIPS 70 and
160$\mu$m photometry.  Finally, we calculate a total uncertainty in
the physical flux density for each photometric measurement by adding
the ``internal'' and calibration uncertainties in quadrature.

Low-resolution (R = 70-120) spectra were obtained on UT 2004 July 13
with the Infrared Spectrograph (IRS: Houck et al.  2004).  We used an
IRS high-accuracy, blue peak-up (1$\sigma$ uncertainty radius =
0.4\arcsec) to acquire the star in the spectrograph slit, thus
minimizing slit losses and assuring high photometric accuracy.  Two
nod positions per cycle were obtained in standard staring mode with
one cycle for the Short-Low wavelength range (7.4-14.5 $\mu$m), and 3
cycles for each of the two Long-Low wavelength ranges (14.0 --
21.3$\mu$m and 19.5 -- 38.0$\mu$m).  The integration times were 6s per
exposure.  The spectrum beyond $\sim 35\mu$m suffers from high noise
(Houck et al.  2004) and has been omitted.

The intermediate {\it droopres} products of the SSC pipeline S11.0.2
were processed within the SMART software package (Higdon et al.
2004).  The background was subtracted using associated pairs of imaged
spectra from the two nodded positions along the slit.  This also
subtracts stray light contamination from the peak-up apertures, and
adjusts pixels with anomalous dark current relative to the reference
dark frames.  Pixels flagged by the SSC pipeline as ``bad'' were
replaced with a value interpolated from an 8 pixel perimeter
surrounding the bad pixel.

The spatially unresolved spectra were extracted using a 6.2 pixel
fixed-width aperture in the spatial dimension for the first order of
the short wavelength low-resolution module, and 5.1 and 3.1 pixels for
the first and second order of the long wavelength low-resolution
module, respectively.  The spectra were calibrated using a spectral
response function derived from IRS spectra and Kurucz stellar models
for a set of 16 stars observed within the FEPS program that exhibit:
1) high signal-to-noise observations, 2) no residual instrumental
artifacts, and 3) no signs of infrared excess.  The absolute flux
density scale was tied to calibrator stars observed by the IRS
instrument team, but reduced as for the rest of the FEPS sample and
referenced to calibrated stellar models provided by the SSC (see also
Hines et al.  2004a, 2005; Bouwman et al.  2005).  Using this
internally consistent approach enables us to improve the absolute flux
calibration and reduces the noise on the adopted relative spectral
response functions.  This ensures that the uncertainties in the final
calibration are dominated by photon noise and not by the uncertainties
in the calibration.  We estimate that the relative flux calibration
across the spectrum is $\sim 1-2\%$.  As for MIPS, a direct comparison
of the IRS spectra for the 33 stars to the predicted flux densities from
our best fit Kurucz models indicates absolute uncertainties of $\approx
6\%$ at 24$\mu$m (\S 3).

Infrared Array Camera (IRAC: Fazio et al.  2004) observations in
channels 1, 2, and 4 (3.6, 4.5, and 8.0$\mu$m) were obtained on UT
2004 July 19 using the 32$\times$32 pixel sub-array mode and a 4-point
random dither pattern, with an effective integration time of 0.01s per
image (frame time = 0.02s).  The 64 images obtained at each of the
four dither positions provided a total of 256 images of each star for
a total integration time of 2.56 seconds in each channel.

All photometry was performed on the Basic Calibrated Data products
from the SSC S10.5.0 data pipeline as described in the SOM4.6 and the
Pipeline Description Document available through the SSC. Aperture
photometry was performed using IDP3 with a 3\farcs6 radius aperture
centered on the target, and the background was estimated by the median
of the pixels in a 12\farcs2 -- 39\arcsec\ radius annulus centered on
the source in each sub-array image.  This annulus circumscribes the
32x32 array and thus uses all pixels outside of a 10-pixel radius from
the target.  The background flux was normalized to the area of the
target aperture and subtracted from the summed target aperture flux.
The reported source flux is the mean of the 256 measures, corrected
from a 3\farcs6 radius to the calibration aperture using values
published in Table 5.7 of the Infrared Array Camera Data Handbook
V1.0.  The ``internal'' uncertainties are estimated from the standard
deviation of the 256 measurements, and are typically $\lesssim $1\%.
The ratios of measured to predicted flux densities for the 33 stars in
the three IRAC bands are 1.08$\pm 0.04$, 1.04$\pm 0.05$, and 0.94$\pm
0.04$ for IRAC 3.6, 4.5 and 8.0$\mu$m, respectively.  This suggests
that there are residual systematic offsets in the absolute calibration
of the IRAC sub-array mode.  These are probably caused by gradients in
the filter band-passes across the detector (Quijada et al.  2004;
Reach et al.  2005), which have yet been fully characterized for the
sub-array mode.  Given these systematic offsets, we conservatively
adopt the 10\% absolute uncertainties as listed for the three bands in
SOM 4.6.


The IRAC and MIPS photometry for HD~12039 is presented in Table~1.  We
also present photometry centered at 13, 24 and 33$\mu$m constructed
from the IRS spectrum using rectangular band-passes of 1.6, 4.7 and
5$\mu$m FWHM respectively.  These IRS flux densities are the
error--weighted means with associated errors in the mean as
appropriate.

\section{The Infrared Excess Associated with HD~12039}


Figure 1 shows a plot of the F$_{\nu}$(24$\mu$m)/F$_{\nu}$(Ks) {\it
vs.} F$_{\nu}$(4.5$\mu$m)/F$_{\nu}$(Ks) flux density ratios for the 33
stars in the sample.  Data are shown from both the MIPS 24$\mu$m band
and the synthetic IRS 24$\mu$m band.  To construct the diagram, we
used our most accurately calibrated IRAC band and the 2MASS $Ks$-band,
which is least affected by extinction.  In addition, since only three
of the stars lie outside of the dust-free Local Bubble ($d > 80$ pc),
and all are within 150 pc, we do not expect extinction in the
$Ks$-band to cause us to identify spurious 24$\mu$m excesses.  These
ratios are independent of model atmosphere fits to the individual
stars, and will identify excess candidates regardless of any relative
calibration offsets between the instruments.

Figure 2 shows the MIPS F$_{\nu}$(24$\mu$m)/F$_{\nu}$(Ks) ratio
plotted versus the IRS F$_{\nu}$(24$\mu$m)/F$_{\nu}$(Ks) ratio
illustrating the independent instrumental calibration for IRS and MIPS
as compared with 2MASS. The 30 Myr old, G3/5V star
HD~12039\footnote{Details concerning our age determination for this
star are given in Appendix~A.} is clearly distinguished in Figures 1
\& 2, and apparently has significant excess emission at 24$\mu$m
detected in both IRS and MIPS independently relative to the other 32
stars.

Since the MIPS and IRS data for each object were obtained $\approx 2$
days apart, the agreement between the flux density measurements for
HD~12039 rules out a chance superposition of an asteroid in one of the
observations as the explanation for the excess.  Furthermore,
inspection of 2MASS images shows that there are no other objects
having $Ks < 15$ mag within 5\arcsec\ of the position of this star,
and the 24$\mu$m image is indistinguishable from a single, isolated
point source.  Therefore the measured 24$\mu$m emission is associated
directly with HD~12039.  The excess in the
F$_{\nu}$(24$\mu$m)/F$_{\nu}$(Ks) ratios are observed at the 4.2 and
4.3$\sigma$ level relative to the median ratios for the other 32 stars
from the MIPS 24$\mu$m and IRS 24$\mu$m photometry (Figs.  1 \& 2).
The probability of drawing this $\ge 4\sigma$ result from a Gaussian
distribution of data for these 33 targets is $< 33 \times
1.3\times10^{-4} = 4\times10^{-3}$.

As a further check on the reality of this excess at 24$\mu$m, we
compared the distributions of of the ratios our measured MIPS and IRS
24$\mu$m flux densities to the flux densities predicted by Kurucz
model atmospheres fit to optical and near-IR photometry for all 33
stars (Figure 3).  Excluding HD~12039, the median and standard
deviations ($1\sigma$) of these distributions are 0.98 $\pm$ 0.05 and
0.99 $\pm$ 0.06 for MIPS 24$\mu$m and IRS 24$\mu$m respectively,
suggesting that the MIPS and IRS absolute flux calibrations are
accurate to 5-6\% relative to the best fit Kurucz models.
The ratio of measured-to-Kurucz model 24$\mu$m flux densities obtained
with both instruments for HD~12309 is 1.33$\pm 0.04$, and [F$_{\nu,\rm
obs}(24\mu$m) - F$_{\nu,\rm model}(24\mu$m)]/$\sigma$ is 5.4 and 5.3
for MIPS and IRS, respectively, where $\sigma$ is the $rss$ of the
standard deviation of the distributions in Figure 3 and the total
measurement uncertainty for HD~12039.  Based upon the above analysis,
we conclude that HD~12039 has excess emission above its stellar
photosphere at 24$\mu$m.

Figure 4 shows the IRAC and MIPS photometry plus the IRS low
resolution spectrum of HD~12039 compared to the model photosphere.
The object was not detected at 70$\mu$m with a 3$\sigma$ upper limit
that is 6 times the expected photospheric flux.  Upper limits are also
shown for 160$\mu$m, 1.2mm, and 3.1mm.

The photospheric emission component of the spectrum in Figure 4 was
fit with Kurucz model atmospheres (including convective overshoot) to
the optical and near-infrared photometry.  Predicted magnitudes were
computed as in Cohen et al.\ (2003 and references therein) using the
combined system response of detector, filter and atmosphere (for
ground-based observations).  The best-fit model was computed in a
least squares sense with the effective temperature and angular
diameter of the star as free parameters, [Fe/H] fixed to solar
metallicity, surface gravity fixed to the value appropriate for the
adopted stellar age and mass (log~g~=~4 in this case), and visual
extinction fixed to 0 mag within the dust-free Local Bubble (d$_{\ast}
\approx 42$ pc).  The derived best-fit stellar model has T$_{eff}$ =
5688K, consistent with HD~12039's catalog G3/5V spectral type.

In comparing {\it Spitzer} data to the model photosphere we have
adopted the weighted average wavelengths that correspond to the
monochromatic flux density within the filter band-passes according to
the SOM4.6. Also, for the purposes of plotting the data, a color
correction ($F_{c} = 1.056\times F_{\rm obs}$) has been applied to the
MIPS 24$\mu$m point assuming a color temperature $T = 110$K estimated
from the slope of the IRS spectrum from 24$\mu$m to 33$\mu$m (see
below)\footnote{MIPS color-corrections are tabulated in the MIPS Data
Handbook V3.0 available from the SSC.}.

The IRAC and short wavelength IRS data for HD~12039 are consistent
with the best-fit Kurucz model.  However, the IRS spectrum begins to
depart from the photosphere between 12-14$\mu$m, passes through the
MIPS 24$\mu$m datum, and continues to increase until at least
33$\mu$m.  This is shown more clearly in the bottom panel of Figure 4,
where we have plotted the ratio of the measured flux densities to the
model-predicted flux densities.  We also show the synthetic 33$\mu$m
photometric point from the IRS spectrum, which lies $\gtrsim
3.5\sigma$ above the Kurucz model.

The lack of significant  excess continuum emission for wavelengths
$\lesssim 14\mu$m and the measured 24$\mu$m excess together imply a
maximum color temperature of $T_{\rm c} \lesssim 160$K. The slope and
uncertainty between the MIPS 24$\mu$m and IRS 33$\mu$m broadband
photometric points correspond to color temperatures $T_{\rm c} =
109^{+48}_{-37}$K, $\sim 110$K. The 70$\mu$m upper limit constrains
the $33-70\mu$m color temperature to $\ge 80$K.

In the next section we further characterize the infrared excess, and
explore circumstellar disk models.  We also consider constraints on
the nature of the thermally emitting material imposed by timescales
for evolution of dust in the system.

\section{Analysis}

\subsection{General Characteristics of the Dusty Debris System}

To further characterize the infrared excess, we assume that the
emission originates from large grains that absorb and emit radiation
efficiently at all relevant wavelengths; that is, we treat the grains
as blackbodies (e.g., Aumann et al.  1984; Walker \& Wolstencroft
1988; Backman \& Paresce 1993).  Blackbody emission at wavelengths as
long as $40\mu$m requires grain radii larger than $a \sim 7\mu$m, and
for such grains the dust physical temperature T$_{\rm dust}$ is the
same as the color temperature.

Given the fitted color temperature of T $\sim 110$K and assuming
blackbody grains in thermal equilibrium with the stellar radiation
field, we calculate a typical grain location of 6 AU from the star.
We also derive an upper limit to the grain radial distance by
considering the 3$\sigma$ upper limit to the 70$\mu$m flux density
(Figs.  3 \& 4).  Only blackbody grain models with R$_{\rm out} <
11$AU do not violate this constraint.  Thus, the observed infrared
excess spectrum for HD~12039 is consistent with a fairly narrow range
of dust temperatures and radial annuli.  We note that although the IR
emission from HD~12039 may be dominated by such relatively large
grains, the characteristic emission temperature of 110K is low enough
that the lack of solid state emission features at e.g.\ 10 and
18$\mu$m from the bending and stretching modes of standard
astronomical silicates does not exclude the presence of smaller
grains.

The observed excess at 24$\mu$m and limit at $\sim 14\mu$m excludes
the presence of material at T $\gtrsim 200$K with more than $\sim
0.1\times$ the surface density inferred for the material near 110K,
assuming that any material at higher temperatures and smaller radii
from the star would have a surface density $\propto r^{0}$ as for
Poynting-Robertson (P-R) drag.  Similarly, the $70\mu$m limit excludes
material at T$\sim 50$K with more than $\sim 0.5\times$ the total
grain cross-sectional area of material at 110K.\footnote{In this case
we can only place limits on dust emitting at a single temperature;
there are no constraints on the surface density.} The fractional
infrared luminosity of the excess across the wavelength range
24--70$\mu$m is f $=$ L$_{\rm IR}$/L$_{\ast} = 1\times10^{-4}$.


\subsection{Disk Models}

We further investigate the HD~12039 debris system using the dust disk
models of Wolf \& Hillenbrand (2003), which take into account
absorption and emission from dust grains based on their optical
properties as opposed to the simple blackbody assumption.  In
particular, we use ``astronomical silicates'' (Draine \& Lee 1984;
Laor \& Draine 1993; Weingartner \& Draine 2001).  We assume a volume
density profile \mbox{$n(r) \propto r^{-1}$} that corresponds to a
disk with constant mass surface density \mbox{$\Sigma(r) \propto
r^0$}, consistent with but not requiring dynamic control by P-R drag
(\S 5.3).

As discussed above, the wavelength at which the dust re-emission
departs significantly from the stellar photosphere ($\lambda \sim
14\,\mu$m) yields an initial estimate of the inner radius of the disk
$R_{\rm in}$.  This is also the location of the warmest material, but
the dust temperature of T$\sim 110$K is too low to produce detectable
silicate emission features (see, e.g., Fig.  10 in Wolf \& Hillenbrand
(2003) and their discussion of the effects of inner holes, thus
maximum temperatures, on the predicted spectra of debris disks).

To explain the spectrum of the infrared excess in HD~12039, we
calculated the inner and outer disk radii and corresponding total dust
mass for a suite of single-size-grain models with particle radii $a =
0.1, 0.5, 1, 3, 7, 25, 50$ and 100$\mu$m.  We applied the
Levenberg-Marquardt algorithm (Marquardt 1963; Markwardt 2003) to find
the models that best fit the observations of HD~12039 in a
chi--squared sense (Rodmann et al.  in prep).  The parameters of these
models are presented in Table 2 and illustrated graphically in Figure
5.  Models assuming small grains require large orbital radii from the
star ($R_{\rm in} \approx$ 10-40 AU, $R_{\rm out} \approx$ 20-50 AU),
because these small grains have low radiative efficiencies compared
with blackbody grains.  For grain sizes $a \ge 7\mu$m, we find values
of \mbox{$R_{\rm in}=4\,$AU} and \mbox{$R_{\rm out}=6\,$AU},
approximately independent of particle size, consistent with the simple
blackbody dust model.  Table 2 and Figure 5 also show that a model
assuming $a = 7\mu$m grains (located between 4-6AU) yields the minimum
dust mass.  These single-size-grain models illustrate the inherent
degeneracies in fitting spectral data alone without independent
morphological information such as direct imaging of the systems in
thermal emission or scattered light.

Real systems in which the dust is produced by collisions of parent
bodies will likely be composed of a distribution of particle sizes
(Dohnanyi 1969; Williams \& Wetherill 1994; Tanaka et al.  1996).
Therefore, we also modeled the infrared excess for HD~12039 using a
power-law grain size distribution \mbox{$n(a) \propto a^{-p}$} that
included small grains of radius \mbox{$a_{\rm min} \approx 0.4\,\mu$m}
(just smaller than the blowout size, see below) and a maximum grain
size of \mbox{$a_{\rm max}=1000\,\mu$m} (just larger than can
efficiently radiate at {\it Spitzer} wavelengths).  The power-law
exponent was set to $p=3.5$, which is produced by a collisional
cascade.  For such a distribution, most of the opacity is supplied by
the smallest grains and most of the mass in is contained in the
largest grains.  For these assumptions the best-fit inner and outer
disk radii are \mbox{$R_{\rm in}=28\,$AU} and \mbox{$R_{\rm
out}=40\,$AU}, respectively (Table 2).  In general, more mass is
required by this model than by the single-size-grain models considered
above because of the inclusion of very large grains with small ratios
of emitting area to mass.

Disk models and the photosphere-subtracted {\it Spitzer} data for
HD~12039 are shown in Figure 6.  All of the models require an inner
hole radius $R_{\rm in}\ge 4$AU, but we are unable to choose between
the various models from the spectral fitting alone.  We therefore
consider, in the next section, physical scenarios for dust production
and migration that lead us to a preferred model.

\subsection{Estimating Time Scales for Grain Evolution}
We further constrain properties of the dust debris system around
HD~12039 by considering the interaction between dust grains and
radiation from the star (e.g., Burns et al.  1979; Backman \& Paresce
1993; Wyatt et al.  1999, and references therein; Th{\' e}bault et al.
2003).  Assuming there is no remnant gas in the system, typical
astrophysical silicates with radius $a < 0.5\mu$m and density $\rho$ =
2.5 g cm$^{-3}$ produced by parent bodies in circular orbits
at any distance surrounding a star with the luminosity and mass of
HD~12039 will be expelled by radiation pressure.
When grains travel a factor of four farther from the star than from
where they originated, their equilibrium temperature falls by a factor
of two and they are no longer important contributors to the observed
infrared emission.  The relevant timescales for this movement are
$\sim 8$yrs and $\sim 80$yrs for grains starting at 6AU and 30AU
respectively.
Ice grains ($\rho \sim 1.0$ g cm$^{-3}$), assumed to be ``dirty" with
albedos similar to silicates, will be similarly removed if $a <
1.5\mu$m.  Therefore, the presence of grains smaller than the blowout
size would signal a very recent
collisional event.

In addition to radiation pressure expelling small grains from the
system, P-R drag will cause inward migration of particles in the
absence of remnant circumstellar gas.  The time scales for silicate
grains ($\rho$ = 2.5 g cm$^{-3}$) to spiral into the star from a
release radius $r_{\rm dust} \sim 6$AU are $t_{P-R} \sim
7\times10^{4}$ or $\sim 5\times10^{5}$ yr for grains with $a \sim
1\mu$m (a bit above the blow-out size) or $a \sim 7\mu$m (the minimum
blackbody emitter size), respectively.  This is much longer than the
blow-out time ($\lesssim 10^{2}$ years), but still short compared to
the $t \sim 30$ Myr age of the system.  Similar sized grains at
30-40AU would have P-R timescales $\sim 1-10$ Myr, again shorter than
the system age.  In either distance regime, small particles are
removed on sufficiently short timescales that the material must be
replenished if we are witnessing a steady-state.  Alternatively, as
mentioned above, the dust could have been produced by a relatively
recent event.

The dust fractional surface density (m$^{2}$ cross-section per m$^{2}$
of disk) near the minimum orbital radius $R_{\rm min}$ is numerically
roughly equal to the fractional luminosity in the excess f =
$L_{IR}/L_\ast$ (Backman 2004),
i.e.\ f$ \sim 1\times10^{-4}$.  From this we can explore the relative
importance of P-R drag compared with grain evolution caused by mutual
collisions (e.g., Wyatt et al.  1999; Wyatt \& Dent 2002; Wyatt 2005).
Considering our two primary disk models discussed above, the dust
mutual collision timescales in this system are approximately
$2\times10^4$ yr for 7$\mu$m grains at $r_{\rm dust} = 6$AU, and
$\approx 2\times10^{5}$ yr for 0.5$\mu$m grains at $r_{\rm dust} =
30$AU with scaling as $r^{1.5}$ (Backman \& Paresce 1993).  At 30AU,
the collision and P-R timescales are comparable, so material moves
towards the star from its region of origin before collisions can
modify it significantly.  At 6AU the collision timescale is 25 times
smaller than the P-R timescale, so collisions can grind material down
to the blowout size before the larger grains migrate towards the star
via P-R drag.  This creates a natural paucity of material inside the
inner radius of the parent body orbits, and could explain the observed
lack of significant material at hotter temperatures (i.e., smaller
orbital radii).  This mechanism for keeping inner holes clear may
operate in other well studied disks (e.g., Wilner et al.  2002;
Dominik \& Decin 2003; Wyatt 2005).

In this analysis we have not considered corpuscular drag from
collisions with the stellar wind from HD~12039, which would increase
the minimum grain blowout size.  Even for young Sun-like stars,
corpuscular drag forces are thought to be small compared with
radiation forces (F$_{\rm wind}$/F$_{\rm P-R} \sim 0.35$: Gustafson
1994).  Very tiny grains (e.g., $a < 0.05\mu$m for silicates) would be
inefficient absorbers at the peak wavelengths of the stellar spectrum
and perhaps stable against radiation blowout, but possibly vulnerable
to ionization and resulting magnetic forces in ways that are poorly
understood.  We can assume that silicate grains with radii at least
over the range $0.05 < a < 0.5\mu$m (or dirty ice grains with $0.1 < a
< 1.5\mu$m) are unstable against ejection from this system.

\section{Discussion}
\subsection{The Debris System}
Our inferred properties of the debris associated with HD~12039 assume
that dust dynamics are dominated by interactions with the stellar
radiation field and are {\it not} determined by dust--gas dynamics.
This requires the gas--to--dust ratio to be $<$ 0.1 (e.g., Takeuchi \&
Artymowicz 2001; Klahr \& Lin 2001).  While this is consistent with
the known properties of HD~12039 (no signatures of active accretion in
the stellar spectrum, no strong mid--IR atomic or molecular
emission--line features), detailed searches for remnant gas have not
been made.  With this caveat, the short grain evolution timescales
relative to the system age imply that the dusty debris consists
primarily of second-generation grains released by collisions and/or
sublimation of larger parent bodies.

We have presented a range of possible models that successfully fit the
thermal excess, but two are of particular interest because they
represent end points of a plausible range of grain physical
characteristics: (1) large ``blackbody" grains ($a \ge 7\mu$m) located
between 4 and 6 AU from the star, and (2) grains with sizes $0.4 \le a
\le 1000\mu$m characterized by a powerlaw distribution, and located
between 28 and 40 AU. Without resolved images of the debris system or
more sensitive searches for mineralogical features, we cannot
distinguish between these models with confidence.  However, the
surface area in model (2) with the wide belt and a grain size
distribution that is dominated by grains near the blowout size would
require either: a) a large mass ($> 100$ M$_{\rm Earth}$) of parent
bodies in collisional equilibrium to maintain the small-grain
population against rapid blowout and P-R drift; or b) the discovery of
a large transient dust signature observable only for a few hundred
years and due to recent collisional event.

Based on planetesimal accretion code models of a minimum-mass solar
nebula (MMSN) surrounding a Sun-like star, Kenyon \& Bromley (2004,
2005) have shown that the growth of planetesimals in the terrestrial
zone can produce $\sim 1$ and $\sim 3$ mag infrared excesses at
10$\mu$m and 24$\mu$m respectively.  The peak emission occurs from
$\sim 10^{4}$ to $\sim 10^{6}$yrs when the largest planetesimals have
reached $\sim 2000$ km in size.  By 10$^{7}$ yrs, the excess is nearly
gone at 10$\mu$m, but persists at 24$\mu$m until $\sim 10^{8}$ yrs.
Importantly, even though the overall dust production (thus IR
emission) declines, individual collisions between 10-100 km bodies
will produce short-lived $\sim 0.5-1.5$ mag signatures (see Fig.  3 in
Kenyon \& Bromley 2004; Kenyon \& Bromley 2005).  The observed $\sim
0.3$ mag excess at 24$\mu$m in HD~12039 is roughly consistent with
this scenario.
In contrast, at 
it is harder to grow large bodies at farther distances from the star.
Collisions at 30 AU are predicted not to be massive enough in general
to produce an observable excess above the photosphere at 24$\mu$m
(Kenyon \& Bromley 2005).
These considerations suggest that our model (1) is the most plausible
explanation for the infrared excess in HD~12039.

The lack of material much warmer than $\sim 110$K in the HD~12039
system could be explained by several mechanisms.  First, this
temperature is approximately equal to the threshold for the onset of
rapid sublimation of micron-sized water ice grains.  Thus, if the dust
parent bodies are primarily icy, the inner edge of the observed debris
distribution could be determined by grain evaporation.  Second, as
discussed in \S5.3, the timescale on which grains collide and produce
fragments small enough for rapid blowout is an order of magnitude
shorter than the P-R timescale at 6 AU, but the timescales are
comparable at 30 AU. This suggests that the inner belt model could be
self-limited.  Dust grains would collide `in situ' before drifting
inward a significant distance from their production location (Wyatt
2005), and the small grains would be blown out by radiation pressure.
The existence of a planetesimal belt surrounding HD~12039 might
require the presence of a nearby giant planet to stop the planet
formation process, as is understood to be the case with Jupiter and
the asteroid belt in our solar system.
Third, a planet could exist at the inner edge of the debris annulus
and consume/scatter inbound grains (e.g., Moro-Mart{\'{\i}}n, Wolf \&
Malhotra 2005).

At the distance of HD~12039, and with L$_{\rm IR}$/L$_{\ast}\approx
1\times10^{-4}$ in the range of objects detected previously in
scattered light with {\it HST} (e.g., Schneider et al.  1999; Ardila
et al.  2004; Fraquelli et al.  2004), the debris disk should be
easily resolved in scattered light by coronagraphic imaging with
NICMOS aboard {\it HST}, if it is composed of small grains at 30AU. A
null detection would strengthen the case that the material is
concentrated in a belt at 4-6AU. NICMOS (and ground-based AO) imaging
should also place limits on the presence of a planet perturber that
may be responsible for stirring up the debris.

\subsection{Terrestrial Planetary Debris Systems}
HD~12039 is the only object among the 33 Sun-like stars in this FEPS
sub-sample that exhibits a prominent debris system at relatively warm
temperatures approaching those of asteroidal material in our own solar
system.  It also apparently lacks significant material at cooler,
Kuiper Belt-like temperatures.  We know of only a handful of other
such ``Terrestrial Planetary Debris Systems'' (Table 3).

Three systems are associated with stars more massive, and hence more
luminous, than the Sun.  The 8 Myr old A0V star HR~4796A shows a
strong infrared excess that is well modeled by a single temperature
blackbody (T = 110K) with grains orbiting at $\sim 35$AU (Jura et al
1998).  Further high resolution imaging of HR4796A in the infrared
indicates the presence of a warmer T = 260 $\pm$ 40 K, compact dust
component located at 4 AU (Wahhaj et al.  2005).  The 10-20 Myr old
F3V star HD~113766A is part of a wide (170 AU) binary that is a
kinematic member of Lower Centaurus Crux association.  The T$\sim
350$K material has been modeled by a debris belt 0.4-6 AU from the
star (Meyer et al.  2001; Chen et al.  2005).  The 300 Myr old A2V
star $\zeta$ Lep was first identified by Aumann \& Probst (1991) to
possess an infrared excess, and Chen \& Jura (2001) find dust of
several hundred K, corresponding to a spatial distribution $<$6-9 AU.

Two additional warm debris systems are associated with stars less
massive than the Sun: HD~98800 and HD~69830.  The 8-10 Myr old
quadruple system HD~98800 in the TW Hya association is comprised of
two spectroscopic binaries with K and M primaries (Torres et al.
1995; Soderblom et al.  1996, 1998).  A circumbinary debris system
first detected by {\it IRAS} (Walker \& Wolstencroft 1988; Zuckerman
\& Becklin 1993) orbits HD~98800B (Koener et al.  2000) is well
characterized by a single T$\approx 160$K blackbody spectrum from
8$\mu$m to 7mm, implying debris orbiting at 2-4AU (Low et al.  1999;
Hines et al.  2004b; Low et al.  2005).  The presence of silicate
emission features suggest that there is an additional population of
small grains in this system (Skinner et al.  1992; Sitko et al.  2000;
Sch{\" u}tz et al.  2005).

The 2 Gyr old K0V star HD~69830 was also identified by {\it IRAS} as
harboring a debris disk system (Mannings \& Barlow 1998).  Similar to
HD~12039, {\it Spitzer} observations of HD~69830 reveal a strong
24$\mu$m excess, and the 70$\mu$m flux density is consistent with the
photosphere (Beichman et al.  2005b).  In contrast to HD~12039,
however, the IRS spectrum of HD~69830 has abundant mineralogical
features similar to those found in {\it ISO} and ground-based mid-IR
spectra of comet Hale-Bopp (Crovisier et al.  1996, 1997; Wooden et
al.  1999a,b).  Beichman et al.  (2005b) model the system with $a =
0.25\mu$m grains at about 3AU, and suggest a recent dust-producing
event.

Finally, the 30Myr old G3/5 star HD~12039 and 300Myr old G0V star
HIP~8920 (Song et a.  2005) each have masses and luminosities very
similar to the Sun.  HIP~8920 is extremely unusual, exhibiting a very
hot ($T_{\rm dust} = 650$K) bright (L$_{\rm IR}$/L$_{\ast} \approx
0.04$) infrared excess.  As for HD~69830, but unlike HD~12039,
HIP~8920 shows a strong 10$\mu$m silicate emission feature signaling
the presence of grains smaller than 3$\mu$m.  Song et al.  (2005) note
these small, hot grains would reside at $\sim 0.4$AU.

Twenty isolated Sun-like stars in the 100 Myr old Pleiades have been
observed by the FEPS program to look for excess infrared emission
(Stauffer et al.  2005).  One ``probable'' and two ``possible''
detections at 24$\mu$m were found.  These qualifiers result from the
fact the Pleiades members are passing through a molecular cloud, and
are embedded in ISM dust that can be illuminated by the stars.  Thus
they may have an associated infrared excess that is unrelated to the
debris disk phenomenon.  Also, the IRS 33$\mu$m and 70$\mu$m upper
limits suggest that these disks, if real, are fairly cool T$\lesssim
85$K (Stauffer et al.  2005).
Therefore, we have not explicitly included these stars in our list of
warm debris disks.

The frequency of warm debris systems associated with Sun-like stars
may be very coarsely estimated based on {\it Spitzer} results
presented to date.  So far, FEPS has found 1/33, Chen et al.  (2005)
find 1/40 (but possibly 3 total), Beichman et al.  (2005a, 2005b) and
Bryden et al.  (2005) have found 1/84.  A MIPS GTO investigation of
the 100Myr old open cluster M47 found 2/17 Sun-like stars with a
24$\mu$m excess, one of which is sufficiently strong that it might be
from terrestrial temperature dust (Gorlova et al.  2004).  Another
{\it Spitzer} investigation of 24 of the known stars within the TW
Hydra Association (TWA: Low et al.  2005) confirmed that HR~4796A and
HD~98800B are the only members of the association to harbor warm
debris disks\footnote{A tentative (2$\sigma$) detection at 12 and
18$\mu$m of a T=170K disk in TWA17 by Weinberger et al.  (2004) was
not confirmed by MIPS 24$\mu$m observations that are consistent with
the stellar photosphere (Low et al.  2005).}.  As discussed above,
FEPS observations of the Pleiades may have also uncovered 3/20 objects
with 24$\mu$m excess.

Compiling these results suggests that approximately $\sim 1-5\%$ of
Sun-like stars in the age range $\sim 8$ Myr -- 300 Myr (plus HD~69830
at 2 Gyr) exhibit infrared excesses peaking at terrestrial
temperatures.
This rarity may reflect the short duty cycle of such
events or, in steady state, an intrinsically low frequency of stars
with the observed level of dust production in terrestrial
planet-building zones.  Kenyon \& Bromley (2005) find that the rate of
collisions sufficiently large to produce a substantial 24$\mu$m signal
is $\le 10^{-5}$ yr$^{-1}$ for the period of terrestrial planet
formation $t\sim 1-100$Myrs\footnote{In our own solar system, the
terrestrial planets are thought to have been 80\% complete by an age
of 30 Myr (Kleine et al 2002, 2003; Yin et al.  2002).}.  Therefore
the apparent paucity of Sun-like stars with bright thermal excesses
and terrestrial temperatures may not be surprising.

\section{Conclusion}

We have discovered a debris system around the Sun-like star HD~12039
that exhibits some properties analogous to the dust associated with
the asteroid belt in our solar system.  While very small particles ($a
\lesssim 1\mu$m) at Kuiper Belt distances ($r = 30$AU) cannot be ruled
out completely, we find that a debris belt consisting of $a \gtrsim
7\mu$m grains orbiting HD~12039 at $r_{\rm dust} = 4-6$AU provides the
most plausible geometry.  As for our solar system, such a narrowly
confined zone of material may signal the presence of (as yet unseen)
planets in the system.

HD~12039 joins only a handful of other debris disk systems with
similar properties out of hundreds observed, suggesting that such
systems are rare.  If this rarity reflects the true incidence of
terrestrial-zone debris, then perhaps asteroid belt analogs are
uncommon.  Alternatively, the collisional events that produce such
strong infrared signals may be rare.


\acknowledgments

It is a pleasure to acknowledge all the members of the FEPS legacy
science team for their contributions to the project.  We thank C.
Beichman for sharing information about HD~69830 prior to publication,
and an anonymous referee whose constructive criticism helped to
improve our presentation.  This work has used the SIMBAD databases,
and is based in part on observations made with the {\it Spitzer Space
Telescope}, which is operated by the Jet Propulsion Laboratory,
California Institute of Technology under NASA contract 1407.  Our
investigation has also made use of data products from the Two Micron
All Sky Survey (2MASS), which is a joint project of the University of
Massachusetts and the Infrared Processing and Analysis
Center/California Institute of Technology, funded by the National
Aeronautics and Space Administration and the National Science
Foundation.  FEPS is pleased to acknowledge support from NASA
contracts 1224768, 1224634, and 1224566 administered through JPL. The
MPIA team is supported by the EU-RTN-HPRN-CT-2002-00308 'PLANETS'
network and the German Research Foundation (DFG) through the Emmy
Noether grant WO 857/2-1.

\appendix
\section{The Age of HD~12039}

Song, Zuckermann \& Bessell (2003) claim that HD~12039 (DK Cet, HIP
9141) is a member of the $\sim$30-Myr-old Tucana-Horologium
association (d$_{\ast} = 42\,\pm$\,2\,pc, Perryman 1997).  We
independently confirmed the kinematic association by comparing the
proper motion (Zacharias et al.  2004) and mean radial velocity
(5.0\,$\pm$\,0.2 km s$^{-1}$; Song et al.  2003; Nordstrom et al.
2004; White et al.  2006) to the values expected for a star at $d$ =
42\,pc with the Tuc-Hor group space motion (Zuckerman \& Webb 2000).

An age of $\sim$30 Myr is also consistent with other estimates.
Position in the HR diagram (T$_{eff}$ =5688$\pm$60K,
log(L/L$_{\odot}$) = -0.05$\pm$0.04 dex: Carpenter et al.  2005)
compared to evolutionary tracks suggests an age of 28 Myr (D'Antona et
al.  1997) to 33 Myr (Baraffe et al.  1998); the mass predicted from
either set of tracks is 1.02M$_{\odot}$.

Chromospheric activity and lithium depletion data corroborate the
isochronal age estimate.  The CaII emission index has been measured to
be log R$^{'}_{HK}$ = -4.21 (Soderblom, private communication) and log
R$^{'}_{HK}$ = -4.14 (White et al.  2006).  This is broadly consistent
with an age of 30 Myr although the uncertainties in estimating ages
from CaII indices for ages $<$ 300 Myr are large.
The \ion{Li}{1} $\lambda$6707 equivalent width (Song et al.  2003;
Wichmann et al.  2003; White et al.  2006) when interpolated among the
distribution of \ion{Li}{1} values for Sun-like cluster members
(Mamajek, in preparation) suggests an age of 15$^{+30}_{-10}$ Myr.
Considering all of these age indicators, we adopt an age of $\sim$30
Myr for HD~12039 based largely on its kinematic membership in the
Tuc--Hor association.

\clearpage



\begin{figure}
\includegraphics[angle=-90,scale=0.8]{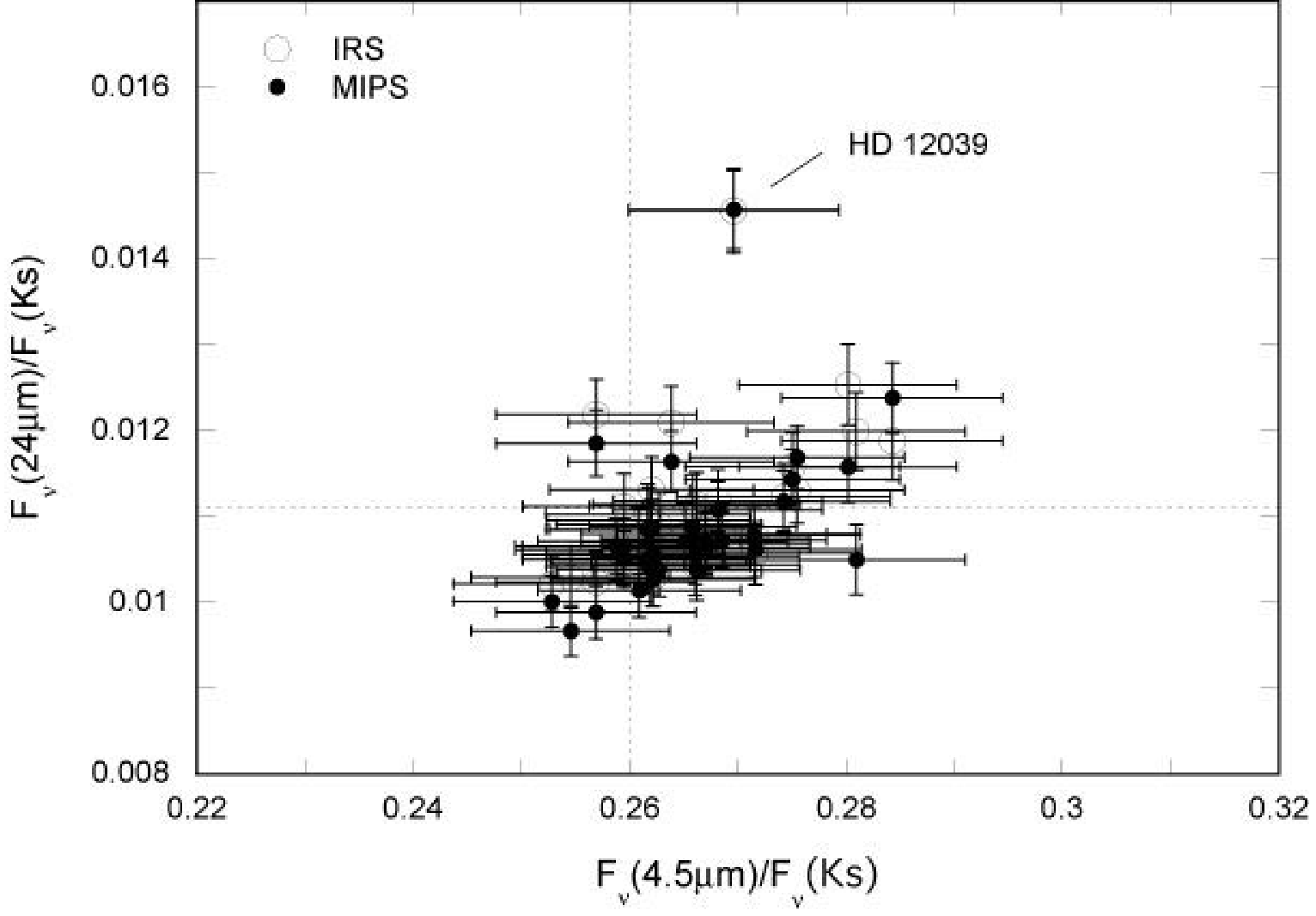}
\caption{\footnotesize Flux ratio diagram
[F$_{\nu}$(24$\mu$m)/F$_{\nu}$(Ks) {\it vs.}
F$_{\nu}$(4.5$\mu$m)/F$_{\nu}$(Ks)] for the 33 stars in the FEPS V1.1
data release.  Data are shown from both the MIPS 24$\mu$m broad band
photometry and from a synthetic broad band point centered at 24$\mu$m
formed from the IRS spectra.  Note that the
F$_{\nu}$(4.5$\mu$m)/F$_{\nu}$(Ks) ratio for HD~12039 is near the
median for the other 32 stars consistent with photospheric colors, but
the F$_{\nu}$(24$\mu$m)/F$_{\nu}$(Ks) ratio is anomalously high
indicating a significant 24$\mu$m excess above the photosphere.  Error
bars represent 1$\sigma$ in the ratios calculated from the internal
IRS and MIPS measurement uncertainties, and the total uncertainties of
the 2MASS K$_{\rm s}$ flux densities.  The 24$\mu$m excess from
HD~12039 is detected in both MIPS and IRS independently and at
different epochs.  The dotted lines indicate the ratios predicted by
the best-fit Kurucz model atmosphere for HD~12039.  Stars with
photospheric temperatures different than HD~12039 will have slightly
different F$_{\nu}$(4.5$\mu$m)/F$_{\nu}$(Ks) ratios, which explains
the dispersion in this quantity for the 33 stars in the sample.  }
\end{figure}

\begin{figure}
\includegraphics[angle=-90,scale=0.8]{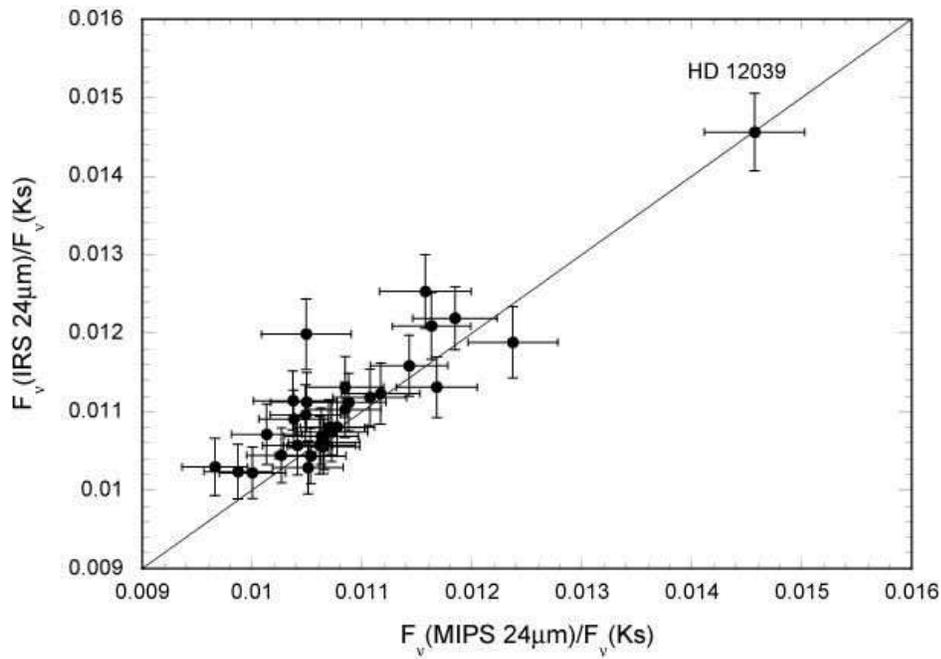}
\caption{The F$_{\nu}$(24$\mu$m)/F$_{\nu}$(Ks) ratio for measurements
from IRS \& MIPS for the 33 stars in the FEPS V1.1 data release.
HD~12039 stands out clearly as having an anomalously high
F$_{\nu}$(24$\mu$m)/F$_{\nu}$(Ks) ratio in both instruments indicating
a significant 24$\mu$m excess above the photosphere.  The solid line is of
slope = 1.  }
\end{figure}

\begin{figure}
\includegraphics[angle=0,scale=0.6]{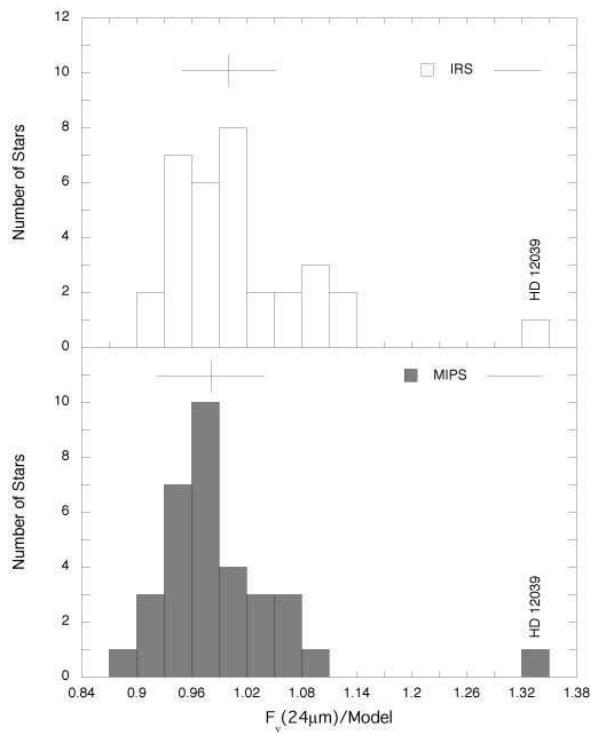}
\caption{Histograms of the ratios of the measured IRS (top) and MIPS
(bottom) 24$\mu$m flux densities to the predicted flux densities from
our best-fit Kurucz models for the 33 stars in the FEPS V1.1 data
release.  The bin width represents the average uncertainty in the
ratios, and the crosses above the histograms indicate the median and
standard deviation of the measurements excluding HD~12039 (1.00 $\pm$
0.06 and 0.98 $\pm$ 0.05 for IRS and MIPS, respectively).  As in
Figures 1 \& 2, HD~12039 stands out clearly as having a significant
24$\mu$m excess above the photosphere, and is detected in both MIPS
and IRS independently and at different epochs.  }
\end{figure}

\begin{figure}
\includegraphics[angle=-90,scale=0.8]{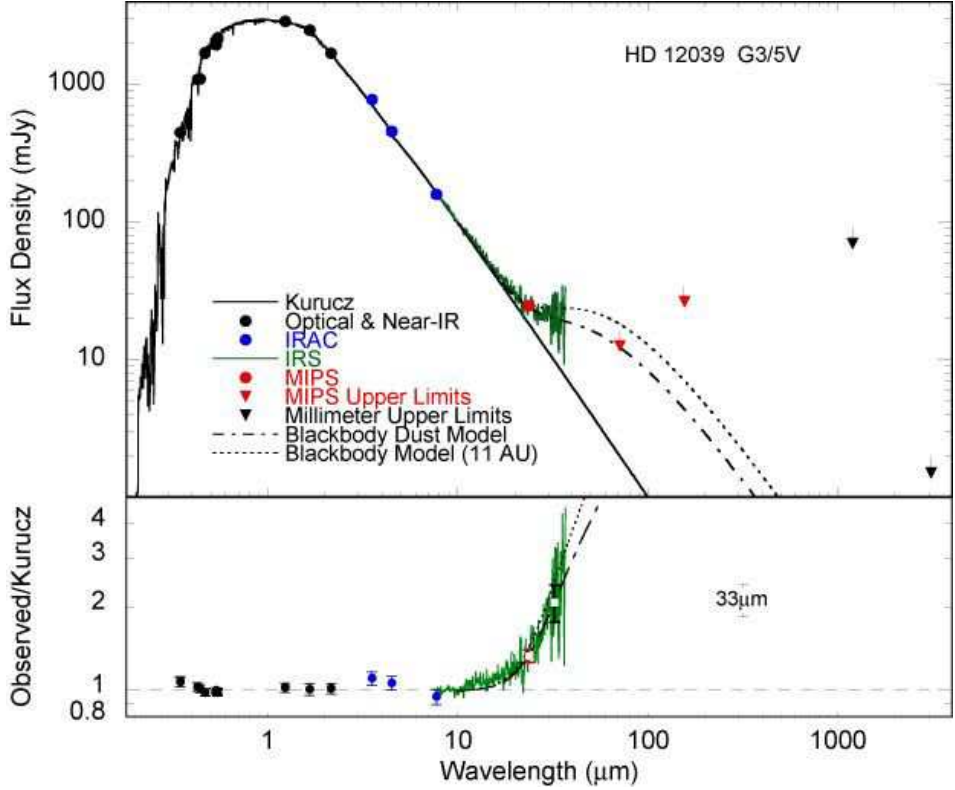}
\caption{Spectrum of HD~12039.  The upper panel shows the ground and
space-based photometry (solid symbols), the IRS spectrum (green
curve), and the Kurucz model that best fits the optical and near-IR
photometry (solid black curve).  The MIPS 24$\mu$m point has been
color-corrected assuming a color temperature T = 110K (see text).  The
1$\sigma$ error bars for each point are plotted, but they are often
smaller than the dot size in the top panel.  We also show upper limits
from MIPS 70 and 160$\mu$m photometry, as well as upper limits at
millimeter wavelengths (Carpenter et al.  2005).  The upper limits
represent the actual on--source measured flux densities minus the
background (which can be negative) plus three times the total
uncertainty including the absolute calibration uncertainty.  Also
shown is the best-fit emission model for simple blackbody grains
(dot-dashed line), and a blackbody model with the outer radius
extending to 11AU (dotted line).  Note that this latter model violates
the 3$\sigma$ upper limit at 70$\mu$m and thus is ruled out by the
observations.  The lower panel shows the spectrum of HD~12039 divided
by the Kurucz model.  The error bars for the IRS 33$\mu$m synthetic
photometric point are duplicated to the right of the actual point for
convenience.  The IR-emission can be seen departing from the
photosphere for $\lambda \gtrsim 12-14\mu$m.  }
\end{figure}


\begin{figure}
\includegraphics[angle=-90,scale=0.7]{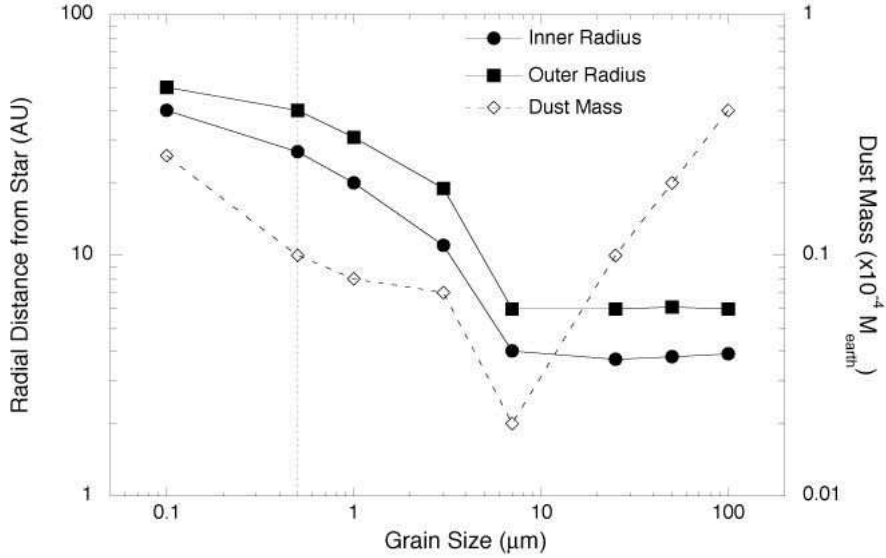}
\caption{The {\it single-size-grain} debris disk models for HD~12039
following the prescription outlined in Wolf \& Hillenbrand (2003).
The models are discussed in \S 5.2 and their parameters are listed in
Table 2.  The inner and outer radial distances of the disk from
HD~12039 and the associated dust mass for each disk model are plotted
as a function of grain size.  The vertical dotted line labels the
blowout size ($a = 0.5\mu$m) for silicate grains; smaller grains are
expelled from the HD~12039 system by radiation pressure on timescales
$\lesssim$ few hundred years.  Note that grains with $a \ge 7\mu$m
behave as blackbodies and orbit in a belt from 4-6AU. The model with
$a = 7\mu$m grains represents the minimum dust mass.}
\end{figure}

\begin{figure}
\includegraphics[angle=90,scale=0.8]{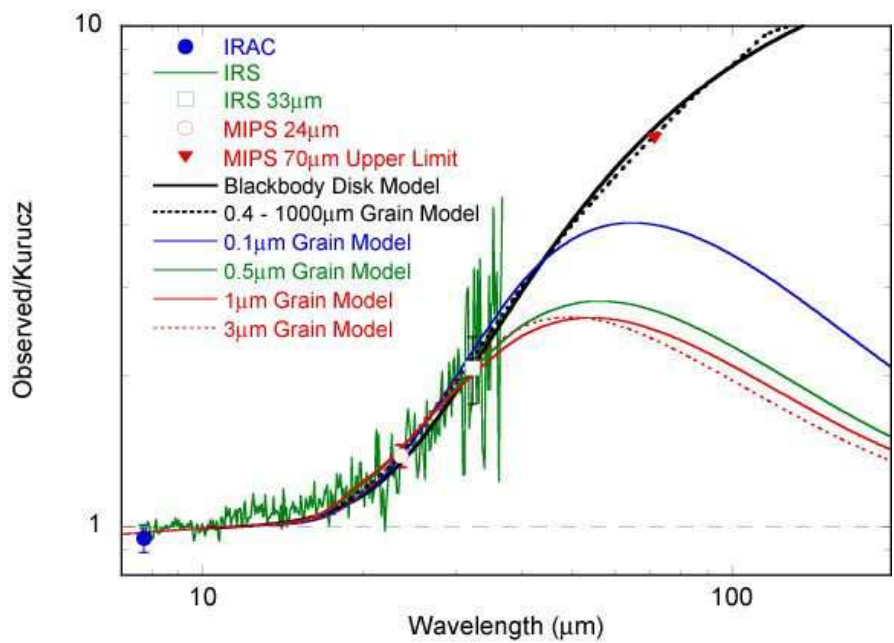}
\caption{The spectrum of HD~12039 compared with spectra produced by
the disk models from Table 2, all divided by the best-fit Kurucz model
atmosphere.  Several single-size-grain models are depicted as well as
a model composed of grains with a powerlaw distribution of sizes from
0.4-1000$\mu$m (Table 2; see text).  The best fit blackbody model is
also shown for comparison.  The single-size-grain dust models with
particle sizes $a \ge 7\mu$m from Table 2 are indistinguishable from
the simple blackbody model, and are not plotted independently in this
figure.}
\end{figure}

\clearpage

\begin{deluxetable}{lccllll}
\tabletypesize{\scriptsize}
\tablecaption{{\it Spitzer} Photometry of HD~12039}
\tablewidth{0pt}
\tablehead{
\colhead{Band} & \colhead{Wavelength\tablenotemark{a}} & 
\colhead{Flux Density} & \colhead{$\sigma_{\rm int}$} & 
\colhead{$\sigma_{\rm tot}$\tablenotemark{b}} & 
\colhead{Flux Density Excess\tablenotemark{c}} & 
\colhead{Instrument}  \\
\colhead{} & \colhead{($\mu$m)} & \colhead{(mJy)} & \colhead{(mJy)} & 
\colhead{(mJy)} & 
\colhead{(mJy)} & 
\colhead{}
}
\startdata

3.6  &  3.535  & 777.50 & 1.43 & 31.10 &  71.51 &  IRAC \\
4.5  &  4.502  & 456.17 & 1.61 & 22.81 &  26.17 &  IRAC \\
8.0  &  7.735  & 158.72 & 1.03 & 6.35  &  -7.28 &  IRAC \\
13\tablenotemark{d}   &  13.17 &  61.65 & 0.16&  9.25 & 4.3 &  IRS \\
24   &  23.68  &  24.65 &  0.22 & 1.50 & 6.1 &  MIPS \\
24\tablenotemark{d,e} & 24.01 & 24.63 &0.38 & 1.53 & 6.0 &IRS \\ 
33\tablenotemark{d}  & 32.36   &  20.31 &  0.23 & 3.05 & 10.5 &  IRS \\
70\tablenotemark{f}  & 71.42    &   4.5 & 2.5 & 2.7 & \ldots & MIPS \\
160\tablenotemark{f} &  155.9   & -42 & 23 &  28 &  \ldots & MIPS \\

\enddata

\tablenotetext{a}{Weighted average wavelengths (SOM 4.6).}

\tablenotetext{b}{The total uncertainty ($\sigma_{\rm tot}$) is the
{\it rss} of the ``internal'' measurement uncertainty (precision) and
the ``calibration'' uncertainties (accuracy).}

\tablenotetext{c}{Calculated relative to the photospheric model
described in the text.}

\tablenotetext{d}{Synthetic band-pass photometry with 1.6, 4.7 and
5$\mu$m band-widths (FWHM) for 13, 24 and 33$\mu$m respectively.}

\tablenotetext{e}{The MIPS 24$\mu$m photometry has not had a 
color-correction applied.  The current value for the color-correction 
would yield $F_{c} = 1.056\times
F_{\rm obs} =$ 25.94 mJy, so the thermal excess above the photosphere 
would be 7.4 mJy).}

\tablenotetext{f}{Upper limits for the purposes of model fitting and
plotting were calculated by F(upper) = F(observed) + 3$\sigma_{\rm
tot}$.}


\end{deluxetable}

\begin{table}[h]
\caption{Model Parameters for the Debris System HD~12039}
\begin{tabular}{c c c c c}
\hline\hline
Model  & Grain radius  & Inner disk  & Outer disk  & Total mass \\
       & ($\mu$m)      & radius (AU) & radius (AU) & $10^{-4}\, {\rm M}_{\rm
earth}$\\
\hline
1a     & 0.1           & 40          & 50          & 0.26 \\
1b     & 0.5           & 27          & 40          & 0.1 \\
1c     & 1             & 20          & 31          & 0.08 \\
1d     & 3             & 11          & 19          & 0.07 \\
1e     & 7             & 4.0         & 6.0         & 0.02 \\
1f     & 25            & 3.7         & 6.0         & 0.1 \\
1g     & 50            & 3.8         & 6.1         & 0.2 \\
1h     & 100           & 3.9         & 6.0         & 0.4 \\           
2      & 0.4\,--\,1000 & 28          & 40          & 4.5 \\

\hline
\end{tabular}
\label{fitpar}
\end{table}

\begin{deluxetable}{lcccccl}
\tabletypesize{\scriptsize}
\tablecaption{Terrestrial Planetary Debris Systems}
\tablewidth{0pt}
\tablehead{ 
\colhead{Star} & \colhead{Sp. Type} & \colhead{age} & 
\colhead{T$_{\rm dust}$} & \colhead{r$_{\rm dust}$}  & 
\colhead{L$_{\rm IR}$/L$_{\ast}$} & \colhead{Reference}\\
\colhead{} & \colhead{} & \colhead{(Myrs)} & 
\colhead{(K)} & \colhead{(AU)}  & 
\colhead{($\times10^{-4}$)} & \colhead{}
}
\startdata
HR 4796A    &  A0V   & 8-10  & 110 & 35    & 50  & 1 \\
$\zeta$ Lep &  A2V   & 300   & 370 & $<$ 6 & 1.7 & 2 \\
HD 113766A  &  F3V   & 15    & 350 & 0.4-6 & 21  & 3 \\											
HIP~8920    &  G0V   & 300   & 650 & 0.4-1 & 400  & 4 \\
HD 12039    &  G3/5V & 30    & 110 & 4-6   & 1   & 5 \\
HD 98800B   & K7V & 8-10 & 160 & 2-4 & 1900 & 6,7 \\
HD 69830    &  K0V   & 2000  & 207 &  1  & 3	  & 8 \\
\hline
Sun         &  G2V   & 4500  & 150-170 & 1-4   & 0.001 & 9,10,11 \\
 
\enddata 
 
\tablerefs{(1) Jura et al.  1993, (2) Chen \& Jura 2001, (3) Chen et
al.  2005, (4) Song et al.  2005, (5) This paper, (6) Low et al.
1999, (7) Koerner et al.  2000, (8) Beichman et al.  2005b, (9) Fixsen
\& Dwek 2002, (10) Reach 1992, (11)  Reach et al.  2003.  }
\end{deluxetable}

\end{document}